\documentstyle[sprocl,epsf]{article}

\def\be{\begin{equation}}
\def\ee{\end{equation}}
\def\bea{\begin{eqnarray}}
\def\eea{\end{eqnarray}}
\def\mintp{\int\!\frac{d^4 p}{(2\pi)^4}}
\def\mintpp{\int\!\frac{d^4 p'}{(2\pi)^4}}

\begin{document}

\rightline{\small UNITUE--THEP--8/98}

\title{Baryons in a Covariant and Confining Diquark--Quark Model$^*$}

\author{G. Hellstern$^\$$,}

\author{M. Oettel, R. Alkofer, H. Reinhardt}
\address{Institute for Theoretical Physics, Auf der Morgenstelle 14,\\ 
72076 T\"ubingen, Germany}


\maketitle\abstracts{
In a covariant model where constituent quarks and diquarks 
interact through quark exchange, the Bethe-Salpeter equation 
in ladder approximation for octet and decuplet baryons
is solved. Quark and diquark confinement is thereby 
effectively parametrised by choosing appropriately 
modified propagators. Numerical results for 
the baryon masses are presented. In a second step 
electromagnetic, weak and pionic currents are 
coupled to the bound state according to Mandelstam's
technique. The arising matrix elements are evaluated in
a generalised impulse approximation and observable 
form factors are extracted.
}
  
\section{Introduction}
With the advent of a new generation of continuous beam facilities
like \mbox{TJNAF}, COSY and  ELSA which investigate baryon observables
at a momentum regime of $Q^2 = 1 - 5\,{\rm GeV}$ to a very high 
precision, a new era of hadron phenomenology 
starts. It is well known, that neither pure 
hadronic models nor non-relativistic quark models are able
to explain the wealth of phenomena arising in the transition region 
between non-perturbative strong interaction physics and the regime where 
QCD perturbation theory sets in. Thus, it proves helpful to look
for effective but also efficient degrees of freedom within a
covariant approach. 

Inspired by studies in the hadronized 
Nambu--Jona-Lasinio model \cite{Rei90} and the Global 
Color model \cite{Cah92}, in two recent publications \cite{Hel97,Oet98}
we worked out a covariant description of baryons, where quarks
and diquarks enter as effective degrees of freedom. In the past,
the notion of diquarks, taking into account some of the unknown features
of non-perturbative QCD, served as a 
valuable tool for various hadron phenomena \cite{Ans93}. 
Apart from being manifestly covariant, our approach includes an effective
parametrisation of quark and diquark confinement.
The framework to examine relativistic bound states is given
by describing the quark exchange between quarks and diquarks within
the Bethe-Salpeter approach in ladder 
approximation. In this way bound state masses and 
wave functions are determined.

\vfill 
\rule{5cm}{.15mm}
\\
\noindent
{\footnotesize $^*$ Supported by BMBF under contract 06TU888 and DFG 
under contract We 1254/4-1.\\
$^\$$ Address after June ${\rm 1}^{st}$ 1998: Deutsche Bank, Frankfurt.}

\pagebreak[4]
\section{Bethe-Salpeter Equation for Baryons}
In the covariant diquark-quark model \cite{Hel97,Oet98},
a baryon is described as a 
bound state of a constituent quark and a scalar ($0^{+}$)
or axialvector ($1^{+}$) diquark interacting through quark 
exchange. The corresponding Bethe-Salpeter equation (BSE)
for octet and decuplet baryons in ladder approximation is given by
\footnote{We use an  Euclidean space formulation
with $\{\gamma_\mu,\gamma_\nu\} = 2 \delta_{\mu \nu}, 
\gamma_\mu^{\dagger}= \gamma_\mu$ and\\
$p \cdot q = \sum_{i=1}^{4} 
= p_i q_i$.},
\bea
\label{sc} 
 \pmatrix{\Psi_8(p;P) \cr \Psi^{\nu}_8(p;P) \cr} &=& - |g_s|^2
  \pmatrix{D(p_b) & 0 \cr 0 & D^{\nu\mu}(p_b) \cr}
  \, S(p_a) \\
  & & 
\!\!\!\!\!\!\!\!\!\!\!\!\!\!\!\!\!\!\!\!\!\!\!\!\!\!\!\!\!\!\!\!\!\!\!\!
\times \int\frac{d^4p^{\prime}}{(2\pi)^4}
 \pmatrix{ \gamma_5 \tilde S(-q) \gamma_5 & -\sqrt{3} \frac{g_a}{g_s} \, 
 \gamma^{\mu'} \tilde S(-q) \gamma_5 \cr
 -\sqrt{3} \frac{g_a^*}{g_s^*} \, \gamma_5 \tilde S(-q) \gamma^{\mu}
 & -\frac{|g_a|^2}{|g_s|^2} \,\gamma^{\mu'} \tilde S(-q) \gamma^{\mu} \cr} 
 \, \pmatrix{\Psi_8(p';P)
  \cr \Psi^{\mu'}_8(p';P) \cr} \nonumber \\
\Psi^{\nu \rho}_{10}(p;P)& =&
- 2 |g_{a}|^{2} S(p_{a}) D^{\nu \mu}(p_{b})  
\int\frac{d^4p^{\prime}}{(2\pi)^4}
\gamma^{\lambda} \tilde S(-q) \gamma^{\mu} 
\Psi^{\lambda \rho}_{10}(p^{\prime};P).
\label{deltaBSE}
\eea
This equation  determines 
the matrix-valued Bethe-Salpeter wave functions of octet baryons
$\Psi_8(p;P)$ and $\Psi^\nu_8(p;P)$ and decuplet baryons
$\Psi^{\lambda \nu}_{10}(p;P)$, respectively,
which are to be projected on positive parity and spin 1/2
or spin 3/2. In flavor space the wave functions 
correspond to pure octet and decuplet states.
They
explicitly depend 
on the total momentum $P$ of the bound state
and the relative momentum $p'$ or $p$ bet\-ween the two constituents.
In eqs.(\ref{sc},\ref{deltaBSE})
the quark propagator is denoted by $S(q)$, the 
scalar diquark propagator by $D(q)$ and the axialvector 
diquark propagator by $D^{\mu \nu}(q)$, see below.

\subsection{Parametrisation of confined constituents}
One of the striking features of QCD 
is the absence of particles
carrying a net color charge in the physical spectrum. 
Accordingly, the decay of a baryon into free quarks,
or in our approach quarks and diquarks, is forbidden.
Note, in a diquark-quark approach, diquarks as constituents 
of color-less baryons necessarily 
form an anti-triplet representation of the color group.

From Dyson-Schwinger studies of QCD, assuming a suitable 
quark-quark interaction \cite{Rob94}, it is known how quark
confinement may be realized:
The quark self-energies appearing in the propagator
\be
S(p)=-i\gamma\!\cdot\! p \sigma_v(p)+\sigma_s(p)
=[i\gamma\!\cdot\! p A(p) +B(p)]^{-1}
\label{quaprop}
\ee
have no poles on the real $p^2-$axis. This feature can be parametrised
by the choice
\be
\sigma_v(p)=\frac{1-\exp(-d(1+\frac{p^2}{m_q^2})}{p^2+m_q^2}
\quad {\rm and}\quad
\sigma_s(p)=m_q \sigma_v(p).
\ee
Obviously the exponentials remove the mass poles, 
however, thereby in\-tro\-duc\-ing an essential singularity at $p^2=-\infty$.

In refs.\cite{Ben96,Hel97a}
it has been shown that confined diquarks can be obtained 
when one uses an appropriate irreducible quark-quark kernel (beyond
ladder approximation) in the diquark Bethe-Salpeter equation.
A useful parametrisation of confined diquarks is now given by 
the choice
\be
D(p)=\frac{-i Z(p)}{p^2+m_s^2},\quad \quad 
D^{\mu \nu}(p)=\frac{-i Z(p)}{p^2+m_a^2}\delta^{\mu \nu},
\ee
where 
\be
Z(p)=1-\exp(-d(1+\frac{p^2}{m_d^2})), \quad 
m_d=m_s \quad \mbox{\rm or} \quad m_a.
\ee
Note, for point-like (tree level) quarks and diquarks the exponentials
are absent. The parameter $d$ basically controls the damping 
of the confining propagators for space-like momenta.
As shown below, this prescription of confinement excludes 
the decay of a baryon into a free quark and diquark.
\\
In order to obtain a finite quark-diquark interaction in momentum 
space and to simulate the intrinsic structure of diquarks
in a crude way, we modify the propagator of the 
exchanged quark in the BSE (\ref{sc},\ref{deltaBSE})
according to (see also ref.\cite{Kus97}) 
\be
S(q) \rightarrow \tilde S(q) =
S(q)\left( \frac{\Lambda^2}{q^2+\Lambda^2}\right).
\label{dpff}
\ee
\subsection{Numerical solution and results}
Before solving the BSE (\ref{sc},\ref{deltaBSE})
we project the wave functions $\Psi_8(p;P)$,$\Psi_8^\nu(p;P)$,
$\Psi_{10}^{\nu \rho}(p;P)$
onto the relevant spin and parity content, restrict ourselves
to solutions with positive energy and decompose the
resulting covariant Lorentz tensors in the Dirac algebra 
\cite{Hel97,Oet98}. Furthermore, the scalar functions 
multiplying each tensor are expanded in Chebyshev polynomials 
of the second kind \cite{Kus97}. In this way, the BSE
is transformed into a system of coupled homogeneous 
integral equations, which can be solved by 
an iterative algorithm \cite{Oet98a}. The eigenvalue
of the integral equation corresponds to the diquark-quark
coupling strength as a function of the bound state 
mass $M$; the eigenvectors are used to reconstruct
the baryon wave functions. In the actual calculation
${\rm SU(3)_{flavor}}$ is broken by the strange quark mass $m_{st}$
and isospin symmetry is assumed.
\begin{table}[t] \centerline{
\begin{tabular}{|l|l|l|l|} \hline
                    & Exp.   & I     & II    \\ \hline
$d$                 &        & 10    & 1     \\
$\Lambda\,\,$ (GeV) &        & 1     & 1     \\
$m_u\,$ (GeV)       &        & 0.5   & 0.5   \\
$m_{st}$ (GeV)      &        & 0.66  & 0.63  \\ 
$\xi$               &        & 1     & 0.73  \\ \hline
$g_a$               &        & 10.35 & 10.89 \\
$g_s$               &        & 11.99 & 10.77 \\ \hline
$M_\Lambda\,$ (GeV) &1.116   & 1.125 & 1.120 \\
$M_\Sigma\,$ (GeV)  &1.193   & 1.163 & 1.154 \\
$M_\Xi\,$ (GeV)     &1.315   & 1.325 & 1.314 \\ \hline
$M_{\Sigma^*}$ (GeV)&1.384   & 1.383 & 1.373 \\ 
$M_{\Xi^*}$ (GeV)   &1.530   & 1.566 & 1.549 \\
$M_\Omega\,$ (GeV)  &1.672   & 1.723 & 1.699 \\ \hline \hline
$\chi^2$            &        &0.0022 &0.0016 \\ \hline
\end{tabular}} 
\caption{\it Octet and decuplet masses for two different parameter 
sets (I),(II). 
\label{baryonmasses}}
\end{table}

In table (\ref{baryonmasses}) we display our results for the octet
and decuplet masses obtained for two different  
parameter sets \cite{Oet98}.
The nucleon and delta mass have been used as input 
to determine the coupling constants $g_a$ and $g_s$
and are therefore not displayed.
Here, $\xi$ is defined by the ratio $\xi= m_{ab}/(m_a+m_b)$,
where $ab$ is the flavor content of a diquark. As can be seen,
it is possible to get an excellent overall fit to the experimental
spectrum, as indicated by the given $\chi^2$.
Note that the octet-decuplet mass differences 
are solely due to the different interaction
in the octet and decuplet channels; the scalar and axialvector 
diquark masses are assumed 
to be equal.   
The main advantage of our approach is, however, that 
we furthermore have access to the relativistic Bethe-Salpeter
wave functions,
which can be used in calculations of form factors and other
non-static observables, see below.

\begin{figure}[t]
\centerline{{\epsfxsize 6cm \epsfbox{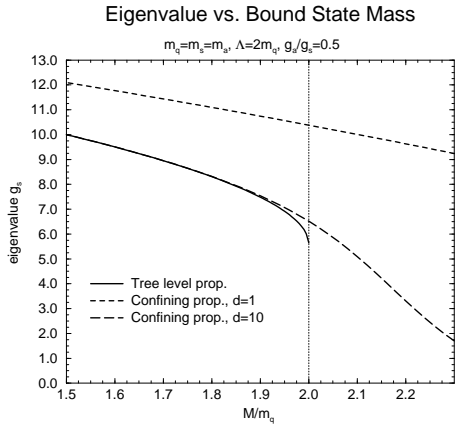}}}
\caption{\it Eigenvalue of the nucleon BSE
versus bound state mass $M$. Threshold 
effects are absent for confining propagators.
\label{thresh}}
\end{figure}

The effect of using confining propagators is clearly seen in 
figure (\ref{thresh}). When using tree level (non-confining)
propagators we are able to determine the eigenvalue of the integral 
equation up to the diquark-quark threshold, here at $M/m_q =2$.
At threshold, the bound state decays into a free
quark and diquark, which is, of course, unphysical.
In case of confining propagators, the threshold vanishes 
and the baryon masses are allowed to be larger than the sum of quark and
diquark masses. 

\section{Nucleon form factors}
In the following we report on a calculation 
of nucleon form factors within the covariant and confining
diquark-quark 
model where only scalar diquarks with $m_s=m_q$ 
are considered \cite{Hel97}.

In the Bethe-Salpeter approach form factors are 
most conveniently 
calculated via Mandelstam's formalism \cite{Man55}:
\be
\langle \hat O \rangle = \frac{1}{2iM} \mintp \mintpp
{\bar \Psi} (p_f;P_f)\Gamma_{\hat O}(p',P';p,P) {\Psi} (p_i;P_i).
\label{MANDELSTAM}
\ee
Besides the normalised BS wave function ${\Psi}$, 
the 5-point function
$\Gamma_{\hat O}$ enters,  
describing  the coupling of 
the probing current to all
internal lines of the diquark-quark system.
 
In the following we use  a generalised impulse approximation,
where only the coupling to the quark and to the scalar 
diquark is considered, 
\be
\Gamma_{\hat O}(p',P';p,P)
=\Gamma^q_{\hat O}(p',P';p,P) + \Gamma^d_{\hat O}(p',P';p,P),
\ee
thereby neglecting the coupling to the exchanged quark.
Using this approximation, the matrix elements split into a quark and 
a diquark part, given by the loop integrals,
containing the fully amputated BS vertex function $\chi(P;p)$,
\be
J_{(\mu)}^q(Q^2)=
\frac{1}{2iM}
\mintp \bar \chi(P_f,p_f)S(p_+)\Gamma_{(\mu)}^{q}(p_+,
p_-)S(p_-)D(p_d)\chi(P_i,p_i)
\label{quarkpart}
\ee
and 
\be
J_{(\mu)}^d(Q^2)=
\frac{1}{2iM}
\mintp \bar \chi(P_f,p_f)D(p_+)\Gamma_{(\mu)}^{d}(p_+,
p_-)D(p_-)S(p_q)\chi(P_i,p_i).
\label{diquarkpart}
\ee
For details see ref.\cite{Hel97}.

\subsection{Electromagnetic form factors}
To calculate electromagnetic (e.m.) form factors, 
the  quark-photon and the diquark-photon vertex functions
entering (\ref{quarkpart}) and (\ref{diquarkpart}) 
have to be specified.
However, their choice can be restricted using gauge symmetry 
of QED, manifest in the Ward and Ward-Takahashi identity.
\\
A longitudinal 
quark-photon vertex function in agreement with gauge symmetry 
is the Ball-Chiu vertex \cite{Bal80},
\bea
\Gamma_\mu^{BC}(p,k)=i \frac{A(p^2)+A(k^2)}{2}\gamma_\mu
+ i \frac{(p+k)_\mu}{p^2-k^2} \{(A(p^2)-A(k^2))\frac{\gamma p + \gamma
k}{2} \nonumber\\*
-i(B(p^2)-B(k^2))\},
\label{BCV}
\eea
which is fully determined by the self-energy functions $A(p^2)$ and
$B(p^2)$ of the quark propagator (\ref{quaprop}).

As has been shown in ref.\cite{Oth90} a longitudinal 
vertex 
function of  a scalar diquark which guarantees gauge invariance 
at the constituent level is given by
\be
\Gamma_\mu^{d}(p,k)=-(p+k)_\mu \frac{D^{-1}(p)-D^{-1}(k)}{p^2-k^2},
\ee
which is fully determined by the self-energy function 
of the diquark propagator. 

In the generalised impulse approximation, the electromagnetic nucleon 
current is denoted  by
\be
J_\mu(Q^2) = Q_q J_\mu^q(Q^2)+ Q_d J_\mu^d(Q^2),
\ee
where the quark and diquark charges, $Q_q$ and $Q_d$, appear.
Using the familiar decomposition of the on-shell e.m. 
nucleon current, where $\Lambda^+$ is the projector 
onto positive energy states,
\be
J_\mu(Q^2)=
\Lambda^+(P_f)[iM_B(F_e(Q^2)-F_m(Q^2))\frac{P_\mu}{P^2}
+F_m(Q^2)\gamma_\mu]
\Lambda^+(P_i),
\nonumber\\*
\ee
the electric and magnetic form factors,
$F_e(Q^2)$ and $F_m(Q^2)$,
can be extracted in a straightforward way.
The left hand side of this equation is calculated 
with the loop integrals (\ref{quarkpart},\ref{diquarkpart}).

\subsection{Pion-nucleon and weak axial form factor}
Considering only scalar diquarks, the coupling of 
an external pion or an axialvector current to the nucleon
is especially
simple: Due to parity these currents solely interact
with the quark but not with the scalar diquark.
Furthermore,
the quark-pion vertex ($=$ pion Bethe-Salpeter amplitude) in the chiral 
limit ($m_\pi=0$) is 
given by the scalar quark self-energy
\be
\Gamma_5^a(P^2=0,p^2)=\gamma_5 \frac{B(p^2)}{f_\pi}\tau^a,
\label{GAMMA5}
\ee
as an immediate consequence of chiral symmetry.
The pion-nucleon
matrix element is commonly parametrised by
\bea
J_5^a(Q^2)
&=&
\Lambda^+(P_f)[\gamma_5 g_{\pi NN}(Q^2)\tau^a]\Lambda^+(P_i),
\eea
including the pion-nucleon form factor  
$g_{\pi NN}(Q^2)$.
The left hand side of this equation is calculated 
with eq. (\ref{quarkpart}) 
supplemented with the quark-pion vertex (\ref{GAMMA5}).

The coupling of an axialvector current to the quark
is implemented via the vertex function
\bea
\Gamma_{5 \mu}^{a}(p,k) &=&
\left(i \frac{A(p^2)+A(k^2)}{2}\gamma_\mu
\right.
\label{ABCV}
\\*
&& 
\hspace{-0.7cm}
+\left.
i\frac{(p+k)_\mu}{p^2-k^2} \left[(A(p^2)-A(k^2))\frac{\gamma p + \gamma k}{2} 
-i(B(p^2)+B(k^2))\right]\right)\gamma_5 \frac{\tau^a}{2} 
,
\nonumber
\eea  
which was constructed to obey the chiral Ward identity\cite{Del79}. 
The invariant functions 
entering this expression are again determined by the 
self-energy functions of the quark propagator.
Using the decomposition 
\bea
J_\mu^a(Q^2)
&=&
\Lambda^+ (P_f)\frac{\tau^a}{2}\left[\gamma_\mu g_A(Q^2)
+Q_\mu g_P(Q^2)\right]\gamma_5 \Lambda^+ (P_i) 
\eea
of the nucleon axialvector current, the form factor $g_A(Q^2)$
can be extracted.

\subsection{Results}
In table (\ref{formff}) we present our results for the 
static form factors ($Q^2=0$) of the nucleon \cite{Hel97} obtained
with tree level and confining  propagators. Due to the 
\pagebreak[4]
%
\begin{figure}[h]
\centerline{{
\epsfxsize 6.5cm
\epsfbox{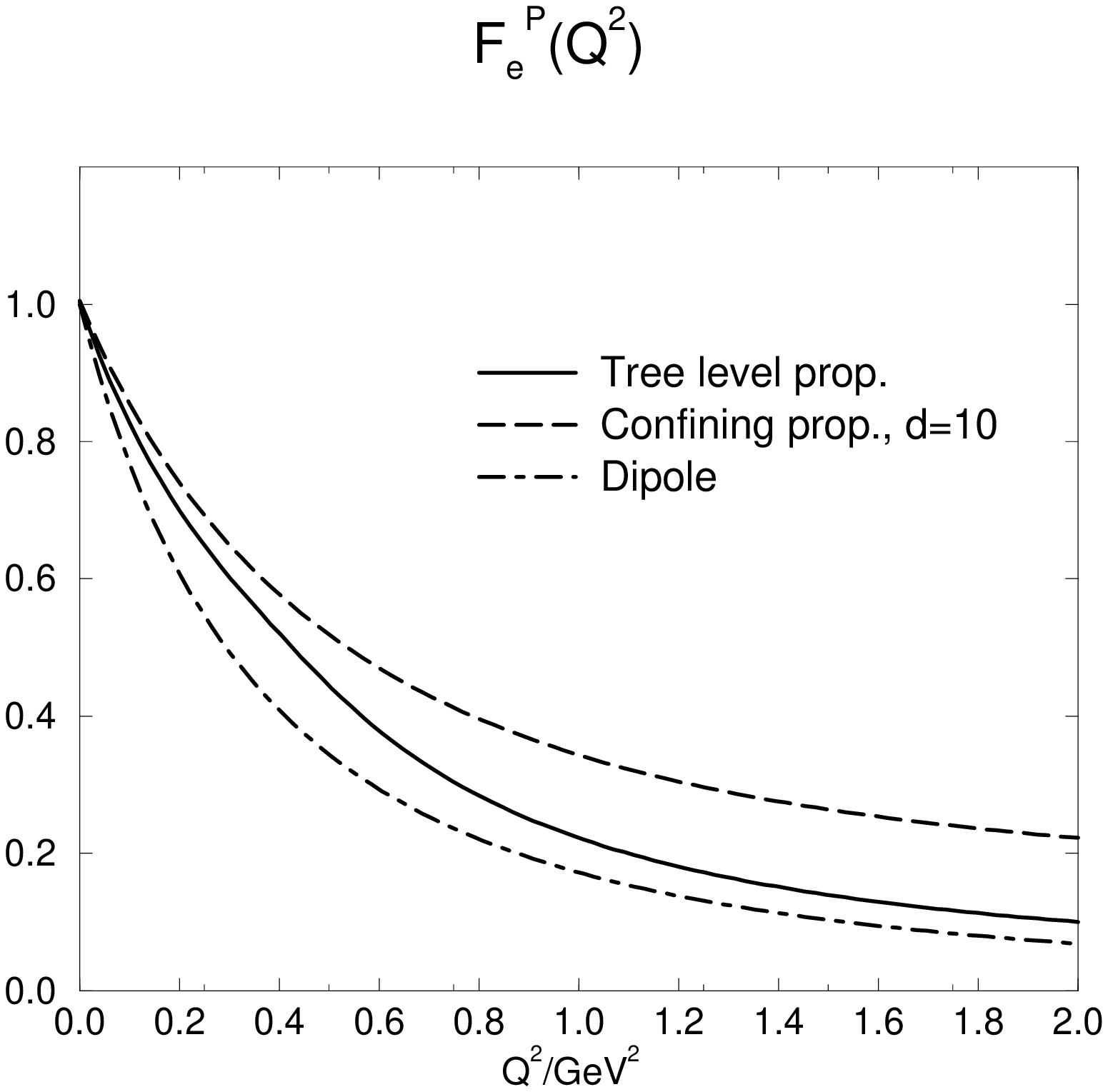}}{
\epsfxsize 6.5cm
\epsfbox{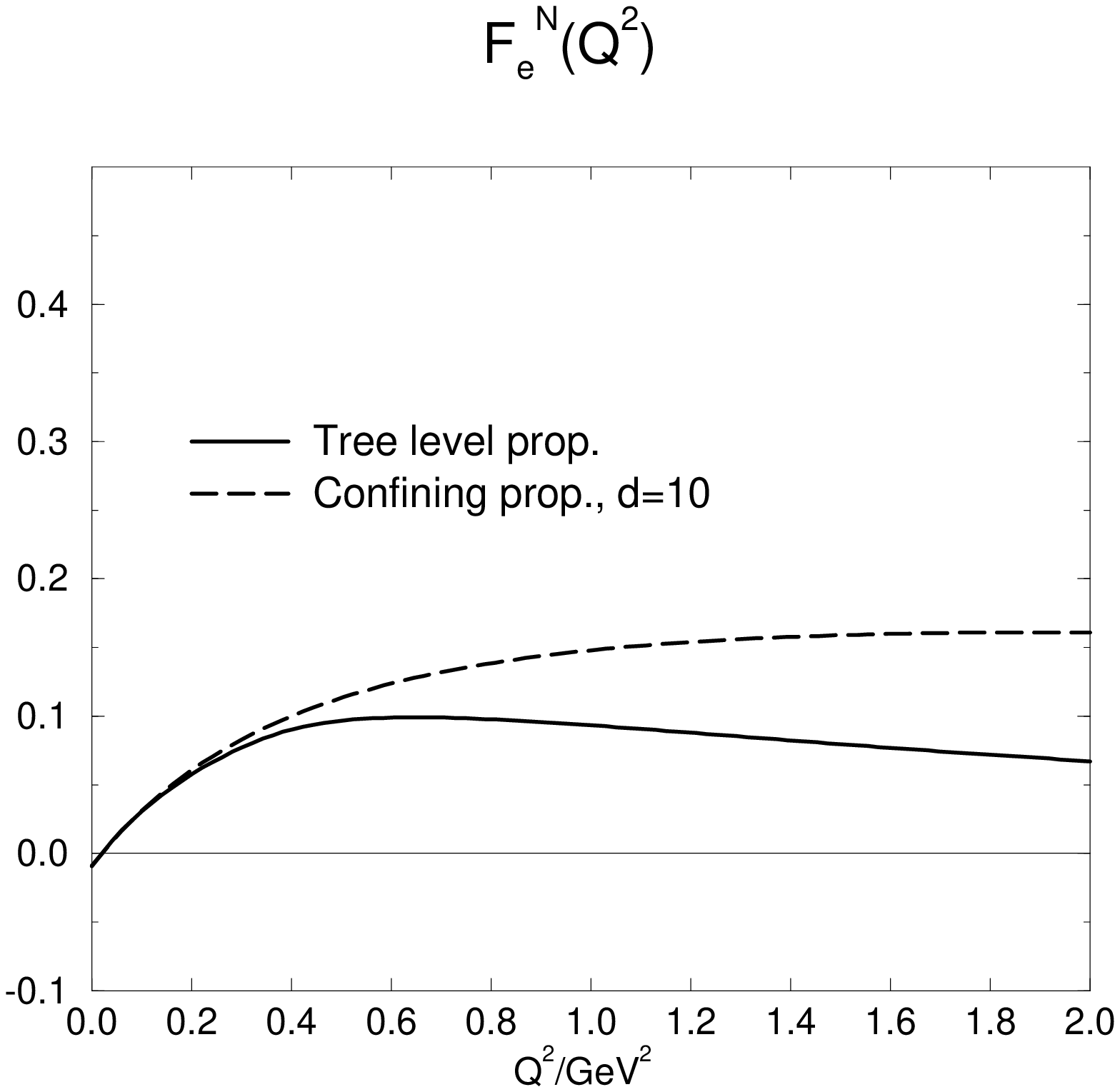}}}
%
\vspace{0.15cm}
%
\centerline{{
\epsfxsize 6.5cm
\epsfbox{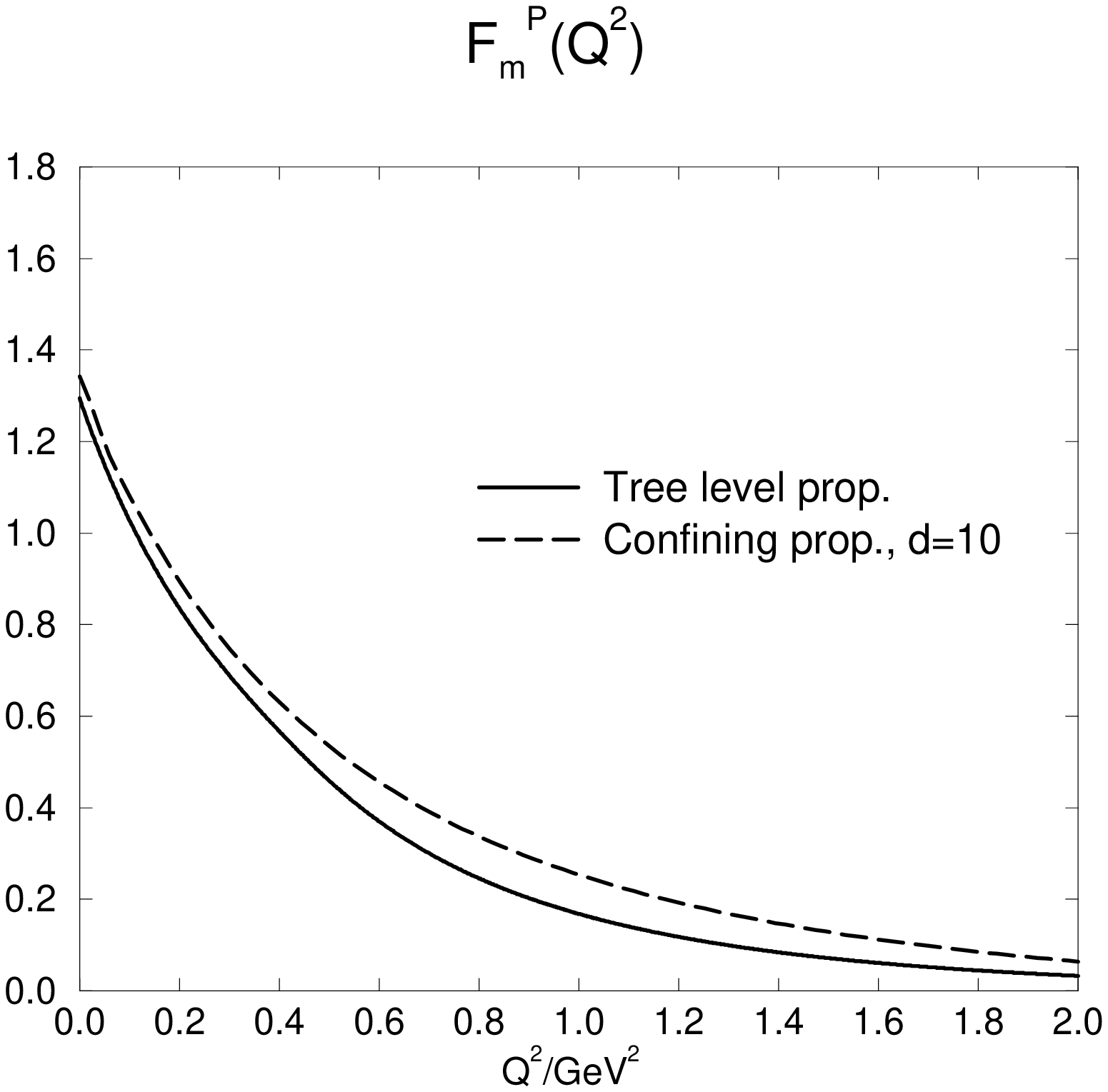}}{
\epsfxsize 6.5cm
\epsfbox{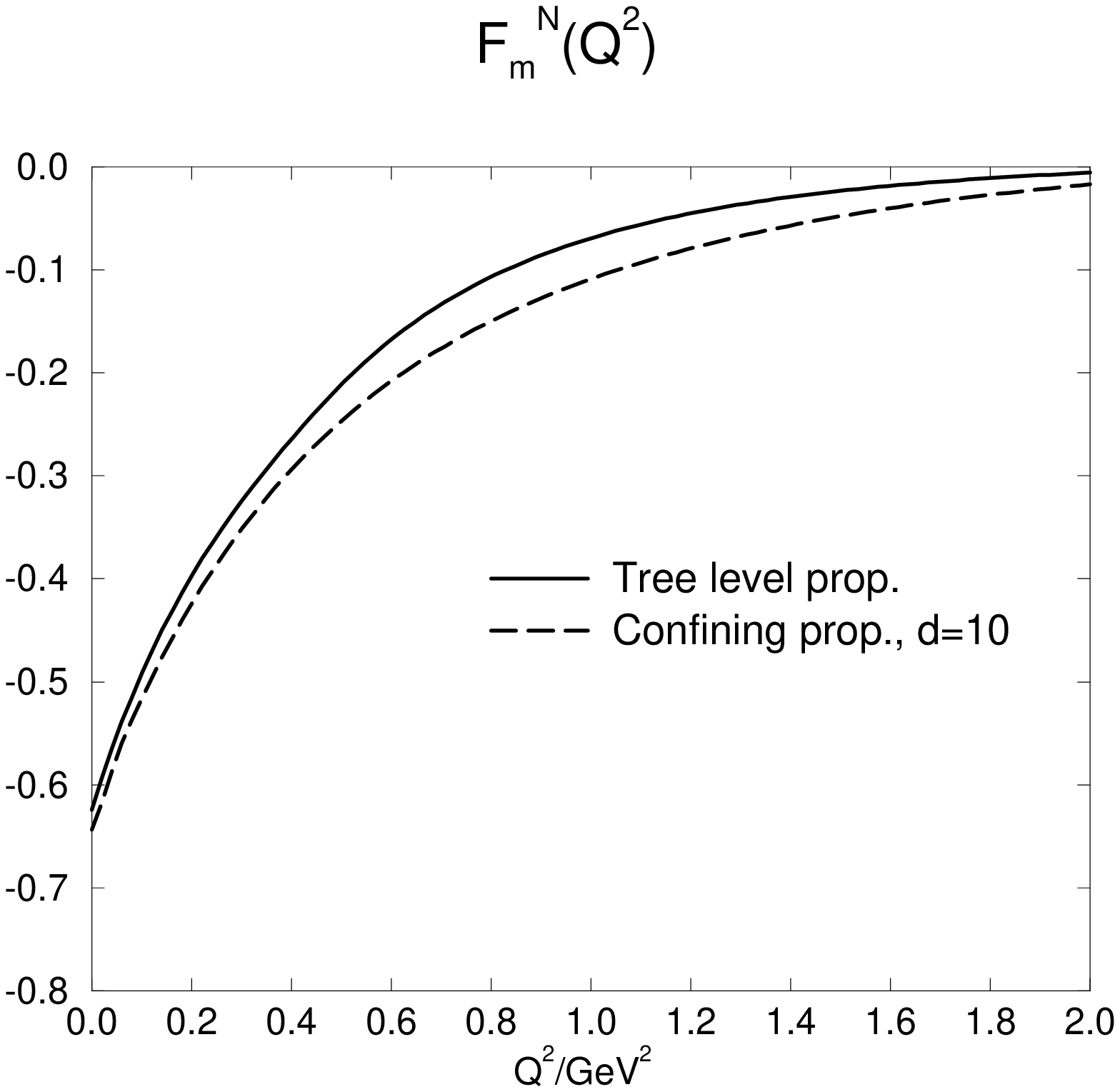}}}
\caption{{\it Electromagnetic nucleon form factors.The nucleon 
mass in all cases is chosen to be
$M=1.9m_q$ and the amplitudes entering the
calculation of the matrix elements are obtained with $\Lambda=2m_q$.
}}
\label{fig2}
\end{figure}
employed generalised impulse approximation, charge conservation
is violated by a few 
percent \footnote{Note, the inclusion of the diagram, where the photon
also couples to the exchanged quark leads to\cite{Sme98}
$Q_P+Q_N = 1$.}. 
Furthermore, the magnetic moments
of proton and neutron are too small, which was also observed 
in similar approaches including only scalar diquarks 
\cite{Hel95,Kei96,Asa96}. The pion-nucleon form factor
and $g_A$, however, are in a reasonable agreement with the 
experimental
values. This indicates the strong coupling of the 
corresponding currents
to the quark due to chiral symmetry. 

The momentum dependence of the e.m. form factors can be seen in figure  
(\ref{fig2}). In the calculation with confining 
propagators and  $d=10$, we get a $Q^2$-behaviour of the form 
factors similar to the calculation with tree level propagators.
However, when using confining propagators we observe 
that the slopes are 
less sensitive to the model parameters and to the effects
of the diquark-quark pseudo-threshold. 
In the figure
showing the electric proton form factor ($F_e^P(Q^2)$)
we also display the empirical dipole fit. It is clearly
seen that our charge radii are smaller than the 
experimental value.
However, in the present stage of the investigation it makes no sense 
to tune the various parameters of the model to get
closer to the dipole curve and to reproduce the 
charge radius of the proton. This has to be done, after 
axialvector diquarks and the coupling
to the exchanged quark are included in the form factor 
calculation. But even without fine-tuning we obtain qualitatively
acceptable results which indicates that the gross features
of baryon structure can be described within our approach.

\begin{table}[t]
\vspace{0.2cm}
\centering
\begin{tabular}{||c||c|c|c||}
\hline
    \multicolumn{4}{||c||}{\rm Nucleon form factors} \\
\hline
        & Tree level prop.  & Conf. prop. , $d=10$   &  Exp.   \\ 
\hline
  $Q_P$ &$1+4.1\cdot 10^{-3}$ &$1+4.7\cdot 10^{-3}$&
 1          \\ 
  $Q_N$ &$-8.2 \cdot 10^{-3}$&$-9.4\cdot 
 10^{-3}$& 0          \\ 
\hline
$\mu_P$ &  1.32        & 1.34              & 2.79   \\ 
$\mu_N$ &  -0.64       & -0.65             & -1.91  \\ 
\hline
$g_A$   &  1.39        & 1.41              & 1.25         \\ 
$g_{\pi NN}$ &  10.59  &  10.89            & 13-14.5 \\ 
\hline
\hline
\end{tabular}
\\
\label{formff}
\caption{{\it 
Numerical results for the static nucleon form factors.
The magnetic moments are given in units of $e/2M$.}}
\end{table}

\section{Conclusion}
In this talk we have presented a covariant diquark-quark 
bound state approach for baryons where the constituents
interact through quark exchange. 
Using an effective parametrisation of confinement we calculated 
with a Bethe-Salpeter equation in ladder approximation 
masses and wave functions of octet and decuplet baryons.
An excellent description of the baryon spectrum is obtained 
with the assumption that 
${\rm SU(3)_{flavor}}$ is broken by the strange quark mass.

As a first application of this baryon model, 
we investigated electromagnetic, pionic 
and axial form factors of the nucleon, considering only 
scalar diquarks.
While for a complete and quantitatively 
reasonable description 
of these observables axialvector diquarks are obviously 
necessary, we nevertheless conclude from our studies, that 
this approach provides a sound basis for 
baryon phenomenology. 

The advantages of 
the covariant and confining diquark-quark model
will be even more visible, when applied to medium
energy reactions. Currently, 
electromagnetic and hadronic 
$\Lambda$-production and 
virtual Compton scattering are therefore 
under investigation.
\section*{Acknowledgements}
G.~H. thanks the organisers of the workshop for the possibility
to present this work. Valuable discussions with A.~Bender
and S.~Brodsky are gratefully acknowledged. 
%

\end{document}